\documentclass[aps,prd,twocolumn,showpacs,preprintnumbers,nofootinbib,10pt]{revtex4-1} 
\pdfoutput=1

\usepackage{graphicx,epsfig}
\usepackage{bm}
\usepackage{latexsym,amssymb,amsmath,float}
\usepackage{subfig}
\usepackage[countmax]{subfloat}
\usepackage{color}
\usepackage[bookmarksnumbered=true,colorlinks=true,linkcolor=black,citecolor=black]{hyperref}

\newcommand{\beq}{\begin{equation}}
\newcommand{\eeq}{\end{equation}}
\newcommand{\bseq}{\addtocounter{subeqno}{1}\begin{subequations}}
\newcommand{\eseq}{\end{subequations}}

\newcommand{\ba}{\begin{array}}
\newcommand{\ea}{\end{array}}
\newcommand{\barr}{\begin{array}}
\newcommand{\earr}{\end{array}}

\def\gs{\mathrel{
   \rlap{\raise 0.511ex \hbox{$>$}}{\lower 0.511ex \hbox{$\sim$}}}}
\def\ls{\mathrel{
   \rlap{\raise 0.511ex \hbox{$<$}}{\lower 0.511ex \hbox{$\sim$}}}}

\def\CO{{\cal O}}

\def\SP{{\sf P}}
\def\SU{{\sf U}}

\newcommand{\e}{{\rm e}}

\renewcommand{\tilde}{\widetilde}

\begin{document}

\title{Neutrino Oscillations and Energy-Momentum Conservation}

\author{HoSeong La}
\email[hsla.avt@gmail.com]{}
\affiliation{Department of Physics and Astronomy,
Vanderbilt University, Nashville, TN 37235, USA }

\date{\today}			

\begin{abstract}
  
\noindent
It is proposed that the energy-momentum expectation values of flavor states
should be identified with the missing energy-momentum of neutrino processes
so that the conservation of energy-momentum can be strictly imposed.
It is also observed that there is a plausible condition to express neutrino mixing angles 
in terms of neutrino masses without invoking symmetries.
  

\end{abstract}

\maketitle

\section{Introduction}
\label{sec:intro}


One of the most mysterious puzzles in the neutrino physics
is the issue of the energy-momentum conservation in the processes involving neutrinos.
The mystery stems from the mismatch between (interacting) flavor eigenstates
and mass eigenstates, even though the concept of neutrinos is born out of respecting
the principle of energy-momentum conservation. 
Normally mass eigenstates, which carry energy-momentum, are supposed to be physically
observable, but not in the neutrino case, in which 
neutrinos are not directly observed. The existence of a 
neutrino is inferred by observing a charged lepton partner with missing energy-momentum. 
In this sense, flavor neutrino states are ``observable."
So, we assume that a neutrino is produced as a coherent linear combination 
of mass eigenstates\footnote{Our practical definition of a coherent state is that it
behaves like a single particle state.}. 
This in turn is also necessary to justify the observed
neutrino oscillations\cite{pdg}. If mass eigenstates are individually produced, we 
might observe mutations of flavor states, but not oscillations.
The trouble now is that flavor states are not energy-momentum eigenstates;
it is believed that 
imposing energy-momentum conservation needs to be done in terms of mass eigenstates.
In the standard derivation of the neutrino oscillations, often a common energy 
or momentum for all mass eigenstates is used\cite{Akhmedov:2009rb}.
This is rather unsatisfactory or could be actually
incorrect\cite{Giunti:2001kj}\cite{Cohen:2008qb}.
The conservation of energy-momentum is one of the most fundamental principles 
in physics, so it should not be circumvented.

Some efforts are made to ensure the energy-momentum conservation. For example,
in \cite{Cohen:2008qb}, an entangled state between simultaneously produced 
particles in the neutrino production process
is constructed as a Hamiltonian eigenstate, whose energy-momentum is just
the sum of those of produced particles as mass eigenstates. So the energy-momentum conservation is respected by the individual mass eigenstate channels
separately.
In this Letter we propose another way of imposing the energy-momentum conservation.
A general quantum state which is a linear combination of Hamiltonian eigenstates
is not an Hamiltonian eigenstate. 
However, it does not mean its energy-momentum expectation value cannot be
computed. We claim that this expectation value should be identified with the 
energy-momentum of a flavor state such that the energy-momentum conservation
can be strictly imposed when this is identified with missing energy-momentum 
of a neutrino process.
This is possible even though flavor neutrino states are not mass eigenstates
because neutrinos are not directly observed.

\section{Energy-Momentum of a General Quantum State}

The folklore of Quantum Mechanics is that only eigenvalues are measurable quantities.
However, this is not a complete story.
Let us consider a quantum state that is a linear combination of 
two different energy-momentum eigenstates
\beq\label{e:1}
|\psi\rangle= c|E_1,\vec{p}_1\rangle+s|E_2,\vec{p}_2\rangle ,
\eeq
where $c^2+s^2=1$. 
We can compute energy-momentum expectation values as
\begin{subequations}
\begin{align}
\label{e:2a}
E &\equiv \langle  \psi| H |\psi\rangle =c^2 E_1 +s^2 E_2 \\
\label{e:2b}
\vec{p} &\equiv \langle  \psi| \vec{p} |\psi\rangle
=c^2 \vec{p}_1 +s^2 \vec{p}_2 .
\end{align}
\end{subequations}
Note that these are not the usual measurable quantities in QM 
since $|\psi\rangle$ is not an eigenstate.
However, here we claim that these expectation values should be counted 
as measurable quantities as well, not just eigenvalues, if a quantum state is 
a coherent linear combination of eigenstates and identifiable, e.g. 
a flavor neutrino state. This is particularly acceptable in the neutrino case
because energy-momentum of a neutrino is never directly measured, but inferred
as missing energy-momentum of a process.
Therefore, even though $|\psi\rangle$ is not an energy-momentum eigenstate, 
we can safely express and call it an effective energy-momentum state as
\beq\label{e:3}
|E,\vec{p}\rangle_{\rm eff} \equiv |\psi\rangle.
\eeq

The difference between an eigenstate and the state like $|\psi\rangle$ 
is in the mass.
We can even define an effective mass of $|\psi\rangle$ as
\beq\label{e:4}
m^2_{\rm eff}(E,\vec{p}) \equiv E^2- p^2 
=c^4 m_1^2 +s^4 m_2^2 +2 c^2 s^2 (E_1 E_2 -\vec{p}_1 \cdot \vec{p}_2).
\eeq
Note that this mass is not constant but dependent on $(E_i,\vec{p}_i)$, which
indicates that this quantum state is not a mass eigenstate.
Nevertheless, it is not wrong to use $(E,\vec{p})$ as energy-momentum 
for such a quantum state.

\section{Energy-Momentum of Neutrino States}

In the neutrino physics, a neutrino is not directly observed, 
neither is its energy-momentum,
but only determined as missing energy-momentum in the process.
Thus it leads to an ambiguity of determining energy-momentum of 
mass eigenstates since they cannot have both identical energy and identical momentum if
their masses are different.
We claim that once this missing energy-momentum is identified 
with the energy-momentum expectation value of a flavor state, 
this ambiguity can be avoided.
This also ensures strict
energy-momentum conservation in the neutrino processes.

In the three-neutrino case (we will drop the ``eff" subscript when it is obvious)
\beq\label{e:6}
|\nu_\ell; E_\ell, \vec{p}_\ell\rangle 
=U_{\ell j} |\nu_j; E_j, \vec{p}_j\rangle,
\eeq
where the matrix $\SU\equiv (U_{\ell j})$ is the PMNS matrix\cite{pdg}.
Then we can compute the energy-momentum expectation values in terms of 
the energy-momentum of mass eigenstates as before such that
\beq\label{e:7}
E_\ell = P_{\ell j}E_j, \quad 
\vec{p}_\ell = P_{\ell j}\vec{p}_j
\eeq
where
\beq\label{e:8}
P_{\ell j}\equiv |U_{\ell j}|^2 ,\quad \SP \equiv (P_{\ell j}).
\eeq

Note that even though flavor states are not mass eigenstates, their energy-momentum 
expectation values do not change as long as flavor states remain to be
coherent states of mass eigenstates. So the energy-momentum expectation
values of all three flavor states can be determined: one at production
and the other two at detection.
Now the energy-momentum of mass eigenstates can be determined 
in terms of determined values of flavor states as
\beq\label{e:10}
E_j =P_{j\ell}^{-1}E_\ell, \quad
\vec{p}_j =P_{j\ell}^{-1}\vec{p}_\ell,
\eeq
where
\beq\label{e:11}
P^{-1}_{j\ell}\equiv (\SP^{-1})_{j\ell} \ \mbox{invertible if}\ \det{\SP}\neq 0.
\eeq
There are the same number of unknowns and knowns except the mixing angles,
so there is no ambiguity and we don't have to make any assumptions on the 
energy-momentum values of mass eigenstates.

The masses now should satisfy
\beq
\label{e:mass}
m_j^2=\left(P_{j\ell}^{-1}E_\ell\right)^2 -\left(P_{j\ell}^{-1}\vec{p}_\ell\right)^2 .
\eeq
Once the PMNS matrix is known, these masses can be uniquely determined in terms of 
measured missing energy-momentum of flavor neutrino states.

\section{Neutrino Oscillations Revisited}

Since we know three mass eigenstates can carry different energy-momentum
(also see \cite{Giunti:2001kj}), we
need to reexamine the neutrino oscillation formula.
A neutrino is produced as a coherent flavor state in a linear combination
of three mass eigenstates. As it propagates, the Hamiltonian evolution 
is to take place according to each mass eigenstate independently.
Thus the evolved coherent state arriving at a detector located 
at distance $L$ will have a form of
\beq\label{e:12}
|\nu_{\ell}; t_j, L\rangle 
= U_{\ell j} \,\e^{i(p_j L -E_j t_j)}|\nu_j; E_j, \vec{p}_j\rangle
\eeq
Contrary to the standard derivation, we use different arrival time $t_j$ 
for different mass eigenstates because each mass eigenstate propagates with 
possibly different speed.
Then the probability amplitude to observe another flavor state is no longer
zero but
\beq\label{e:13}
\langle \nu_{\ell'} |\nu_{\ell}; t_j, L\rangle
=\sum_j U^*_{\ell' j} U_{\ell j}\,\e^{i(p_j L -E_j t_j)}.
\eeq
Since time $t_j$  cannot be directly measured, we should eliminate so that
$L=v_j t_j$ will lead to
\beq\label{e:14}
p_j L-E_j t_j = -{m_j^2 L\over E_j}\left(1+\CO\left(m_j^2/E_j^2\right)\right).
\eeq
Now neutrino oscillation probability is
\beq\label{e:15}
P(\nu_\ell\to\nu_{\ell'})
=\sum_{j,j'} U_{\ell' j'} U^*_{\ell j'} U^*_{\ell' j} U_{\ell j}
\,\exp\left\{{iL\left({m_{j'}^2\over E_{j'}}-{m_j^2\over E_j}\right)}\right\}.
\eeq
The exponential part can be easily turned into the standard one 
by simply rescaling energies and masses as
\beq
\exp\left(iL{\Delta \tilde{m}_{j'j}^2\over 2E}\right), 
\eeq
where
\beq
\label{e:sc}
2E = \alpha_j E_j, \quad \widetilde{m}_j^2=\alpha_j m_j^2.
\eeq
So, in this sense our oscillation formula is equivalent to the standard one. 

In the standard derivation, the factor $2$ in front on $E$ appears 
due to setting the arrival times for all mass eigenstates the same.
However, as was argued in \cite{Field:2002gg}\cite{Kobach:2017osm}, 
this factor is ambiguous,
which leads to the ambiguity of experimental neutrino masses.
(Also see \cite{Bilenky:2005ei} to test this factor.)
In our case, in principle there is no such ambiguity of determining masses
(see eq.(\ref{e:mass}).

\section{Mixing Angles in terms of Masses}

Having introduced a method to impose strict energy-momentum conservation, 
we can observe another interesting outcome. 
Let us first consider two-$\nu$ case as warm-up.
In this case, two flavor states are given in terms of two mass eigenstates 
($m_l < m_h$) as
\beq
\label{e:a1}
\left({\nu_\ell\atop \nu_{\ell'}}\right ) 
= 
\left(
\begin{array}{cc}
c & s \\
-s & c
\end{array}
\right)
\left({\nu_l\atop \nu_h}\right),
\eeq
then
\beq\label{e:a3} 
\mathbf{\SP}\equiv
\left(
\begin{array}{cc}
c^2 & s^2 \\
s^2 & c^2
\end{array}
\right).
\eeq
The effective masses, eq.(\ref{e:4}), are given by
\begin{subequations}
\begin{align}
m_{{\rm eff},\ell}^2
&=\mu_\ell +2 c^2 s^2 (E_l E_h -\vec{p}_l \cdot \vec{p}_h), \\
m_{{\rm eff},\ell'}^2
&=\mu_{\ell'} +2 c^2 s^2 (E_l E_h -\vec{p}_l \cdot \vec{p}_h), 
\end{align}
\end{subequations}
where
\begin{subequations}
\begin{align}
\mu_\ell &\equiv c^4 m_l^2 +s^4 m_h^2 ,\\
\mu_{\ell'} &\equiv s^4 m_l^2 +c^4 m_h^2 ,
\end{align}
\end{subequations}
The behavior of $m_{{\rm eff},\ell}^2$ w.r.t. $x\equiv E_h/E_l$ is interesting.
Assuming $\vec{p}_l$ and $\vec{p}_h$ are colinear, $m_{{\rm eff},\ell}^2$
minimizes at $x\equiv E_h/E_l=m_h/m_l$ and asymptotically approaches 
a straight line as $x\equiv E_h/E_l\to \infty$. This asymptotic trajectory
intercepts at $\mu_\ell$ hypothetically extending to the $x\to 0$ limit.
The interesting fact is that 
$\mu_\ell$ minimizes w.r.t. $s^2$ at
\beq
\label{e:minc}
s^2={m_l^2 \over m_h^2 + m_l^2}
\eeq
such that
\beq
\mu_{\ell,{\rm min}}={m_h^2 m_l^2\over m_h^2+m_l^2}.
\eeq

One could ask what about minimizing $\mu_{\ell'}$. In conclusion, it is redundant.
$\mu_{\ell'}$ minimizes at $s^2 = m_h^2/(m_h^2 +m_l^2)$, which is different from
eq.(\ref{e:minc}) such that $\mu_\ell$ and $\mu_{\ell'}$ do not minimize 
under the same condition. 
At this $s^2$ value, $\mu_{\ell',{\rm min}}=\mu_{\ell,{\rm min}}$.
Therefore, we only need to minimize one of them. 
Which one to choose to minimize? It is just a convention, the other one is redundant. 
If $m_l=m_h$, $s=c$ such that $\det\SP=0$. For our assumption $m_l<m_h$,
i.e. $s<c$, we choose to minimize $\mu_\ell$ 
such that $s^2 <1/2$ and $\mu_\ell <\mu_{\ell'}$.
With this choice, in the $m_h/m_l\to \infty$ limit, eq.(\ref{e:minc}) is nothing but 
the seesaw condition such that $m_{{\rm eff},\ell}$ is of the order of $m_l$ and
$m_{{\rm eff},\ell'}$ of $m_h$.

This observation motivates us to expect that in the three-$\nu$ case
the mixing angles also may be expressed in terms of masses.
Using the standard notation in \cite{pdg}(for argument sake, we neglect the
CP-phase), we can compute the effective masses as
\begin{subequations}
\begin{align}
m^2_{\rm{eff},\ell} &= (P_{\ell j} E_j)^2 -(P_{\ell j} \vec{p}_j)^2\\
&=\mu_\ell 
+\sum_{j\neq j'}P_{\ell j}P_{\ell j'}\left(E_j E_{j'}-\vec{p}_j\cdot\vec{p}_{j'}\right),
\end{align}
\end{subequations}
where
\begin{subequations}
\begin{align}
\mu_e &=c_{12}^4 c_{13}^4 m_1^2 +s_{12}^4 c_{13}^4 m_2^2 +s_{13}^4 m_3^2 , \\
\mu_\mu &=\left(s_{12}c_{23} +c_{12}s_{23}s_{13}\right)^4 m_1^2
+\left(c_{12}c_{23} -s_{12}s_{23}s_{13}\right)^4 m_2^2 \nonumber\\
&\ \ +s_{23}^4 c_{13}^4 m_3^2  ,\\
\mu_\tau &=\left(s_{12}s_{23} -c_{12}c_{23}s_{13}\right)^4 m_1^2
+\left(c_{12}s_{23} +s_{12}c_{23}s_{13}\right)^4 m_2^2 \nonumber\\
&\ \ +c_{23}^4 c_{13}^4 m_3^2
\end{align}
\end{subequations}
As before in the two-neutrino case, we will minimize $\mu_\ell$'s. Note that
$\mu_\mu$ and $\mu_\tau$ are related by swapping $s_{23}$ and $c_{23}$ as well as
flipping the sign of $s_{13}$, so they are
equivalent for our purpose. We just need to minimize $\mu_e$ and $\mu_\mu$.

Minimizing $\mu_e$ w.r.t. $s_{12}^2$ leads to
\beq
\label{e:s12}
s_{12}^2 ={m_1^2 \over m_1^2 +m_2^2}.
\eeq
This is just reminiscence of the two-$\nu$ case, which enables us to choose
$m_2>m_1$.
Minimizing $\mu_e$ w.r.t. $s_{13}^2$ leads to
\beq
\label{e:s13}
s_{13}^2 ={c_{12}^4 m_1^2+s_{12}^4 m_2^2\over c_{12}^4 m_1^2+s_{12}^4 m_2^2 +m_3^2}
={m_1^2 m_2^2 \over m^4},
\eeq
where eq.(\ref{e:s12}) is used for the latter equality and
\beq
m^4 \equiv m_1^2 m_2^2 +m_2^2 m_3^2 +m_3^2 m_1^2.
\eeq

Minimizing $\mu_\mu$ is somewhat involved. Let's first look at the behavior 
w.r.t. $s_{23}$, which is missing from the $\mu_e$ case.
It leads to a cubic equation with both positive and negative roots; the most 
positive root is what we need because we demand $s_{23}>0$\cite{Latimer:2004hd}.
In addition, $s_{13}>0$ by convention, 
the minimum of $\mu_\mu$ we look for is at 
\beq
s_{23}^2 = {m_2^4 \over (m_1^2+m_2^2)(m_2^2 +m_3^2)},
\eeq
provided
\beq
\label{e:xi23}
m_3^2 > {m_1^2 m_2^2 \over 2m_2^2 -m_1^2}.
\eeq
This can happen for both normal hierarchy, i.e. $m_3 > m_2>m_1$, and 
inverted hierarchy, i.e. $m_2>m_1>m_3$.
If $m_3$ does not satisfy eq.(\ref{e:xi23}),
which happens only for the inverted hierarchy, then
\beq
s_{23}^2 ={(m_2^2-m_1^2)^2 m^4 \over
(m_2^2-m_1^2)^2 m^4 
+ \left(3m_1^2 m_2^2 -m^4\right)^2}
\eeq
Note that $s^2_{23}\propto (m_2^2-m_1^2)^2\to 0$ as $m_1 \to m_2$, 
less likely but not entirely rule out.

We have chosen to use a local minimum of $\mu_\mu$
w.r.t. $s_{23}$ because $\mu_e$ is minimized w.r.t. $s_{12}$ and $s_{13}$ such that
$\mu_\mu$'s behavior w.r.t. $s_{12}$ and $s_{13}$ is expected to lead to redundant
conditions as in the two-$\nu$ cases.

To compare to the measured values we need to use eq.(\ref{e:sc}) such that
these mixing angles need to be expressed in terms of experimental
values of masses, $\tilde{m}_j$. This will determine the scale factors
$\alpha_j$ in terms of measured mixing angles and $\tilde{m}_j$,
and that unambiguous neutrino masses can be determined.

\section{Final Remarks}

We have demonstrated that flavor neutrino states, not just mass eigenstates, 
can carry energy-momentum, and that strict energy-momentum conservation 
can be imposed on the neutrino processes in terms of this.
Of course, this does not mean that propagators can be defined for flavor states 
because the effective masses are not constants.
It is also noted that experimentally determined neutrino masses are ambiguous 
since they depend on the neutrino oscillation formula. In our case, in principle 
this ambiguity can be avoided. Once mixing angles are determined experimentally even
using the standard oscillation formula, masses, $m_j$, can be determined by the 
specific relationship to the energy-momentum of flavor states.
The only experimental difficulty is to determine energy-momentum of all three flavor
states from the same beamline. 

We have also observed an interesting byproduct based on the effective masses
of flavor states we have defined: neutrino mixing angles may be 
expressed in terms of masses without invoking symmetries.
Of course, it will be interesting to check if there is any associated symmetries
to make our conjecture more reasonable.
One caveat in our case is that the masses relevant for symmetries are $m_j$,
not $\tilde{m}_j$,  because $m_j$  should be the neutrino masses in the lagrangian.
If we accept our expression of mixing angles in terms of $m_j$, we can avoid
the aforementioned experimental difficulty.
We can just use the standard oscillation data once completed, 
which will determine $s_{ij}$ and $\tilde{m}_j$. They can be in turn used to 
determine $\alpha_j$ and that the real neutrino masses $m_j$.

It is generally believed that at sufficiently long distance mass eigenstates,
traveling at different speeds, would begin to fall apart enough such that
they are no longer coherent to form a flavor state\cite{{Nussinov:1976uw}}. 
This type of phenomena 
is called the decoherence\cite{Zeh:1970zz}\cite{Zeh:2005ja}.
Detected flavor states now will inherit the energy-momentum
of the corresponding mass eigenstate only so that the detected energy-momentum of a flavor state is not necessarily the same as the produced one.
One may wonder if this violates energy-momentum conservation since the measured
final state does not match the missing energy-momentum of the neutrino production.
Note that this decoherence in the neutrino case is 
an irreversible non-unitary process.
So it does not violate the energy-momentum conservation in that.

In this Letter, We have neglected the CP-violating phase, but it is straightforward
to include that: $\mu_\mu$ and $\mu_\tau$ will depend on the phase.
Note that we have considered only  energy-momentum conservation in vacuum.
Nevertheless, we expect that the same idea can be applied to the cases with neutrino
processes in matter. The details will be presented elsewhere.

\bigskip
\noindent
{\bf Acknowledgements:}

The author thanks David Ernst for his interests in this work and bringing 
\cite{Latimer:2004hd} to his attention. He also thanks Tom Weiler for conversations
on related issues.


\end{document}